# Specific Absorbed Fractions of Electrons and Photons for Rad-HUMAN Phantom Using Monte Carlo Method

WANG Wen（王文）[1, 2], CHENG Meng-yun（程梦云）[2], LONG Peng-cheng（龙鹏程）[2], HU Li-qin（胡丽琴）[1, 2*]

[1]*University of Science and Technology of China, Hefei, Anhui, 230027, China*
[2]*Laboratory of Neutronics and Radiation Safety, Institute of Nuclear Energy Safety Technology, Chinese Academy of Sciences, Hefei, Anhui, 230031, China*

**Abstract:** The specific absorbed fractions (SAF) for self- and cross-irradiation are effective tools for the internal dose estimation of inhalation and ingestion intakes of radionuclides. A set of SAFs of photon and electron were calculated using the Rad-HUMAN phantom, a computational voxel phantom of Chinese adult female and created using the color photographic image of the Chinese Visible Human (CVH) data set. The model can represent most of Chinese adult female anatomical characteristics and can be taken as an individual phantom to investigate the difference of internal dose with Caucasians. In this study, the emission of mono-energetic photons and electrons of 10keV to 4MeV energy were calculated using the Monte Carlo particle transport calculation code MCNP. Results were compared with the values from ICRP reference and ORNL models. The results showed that SAF from Rad-HUMAN have the similar trends but larger than those from the other two models. The differences were due to the racial and anatomical differences in organ mass and inter-organ distance. The SAFs based on the Rad-HUMAN phantom provide an accurate and reliable data for internal radiation dose calculations for Chinese female.

## 1. Introduction

Absorbed fractions (AFs) and specific absorbed fractions (SAFs) that account for the partial deposition of radiation energy in target organs and tissues are essential for the calculation of radiation dose of the intakes of radionuclides or a nuclear medicine procedure. The Medical Internal Radiation Dosimetry (MIRD) Committee provided a systematic approach of internal dose calculation which is known as the MIRD schema [1,2]. To obtain the SAFs, AFs and other conversion coefficients, computational phantoms and the Monte Carlo methods are often used. Photon SAFs were calculated using MIRD-type mathematical anthropomorphic phantoms of different ages. The mathematical models were first designed by Fisher and Snyder from Oak Ridge National Laboratory (ORNL) in 1969[3] and were revised in 1978[4]. They were adapted by the Medical Internal Radiation Dose (MIRD) Committee as the MIRD-type models, and have been evolved into several improved and extended versions for dosimetry calculation [5,6,7]. The MIRD models facilitate the rapid dose calculation but suffer from the loss of most anatomic details.

To perform accurate absorbed-dose calculation, voxel models were developed based on computed tomography (CT), magnetic resonance imaging (MRI) or colored photographs which can provide more realistic and detailed information of the human anatomy. The second generation phantom, using medical image data of real human body, is now used for internal dosimetry and can significantly contribute to better dose assessment. Nowadays, many individual voxel-based models have been reported in the world. The first voxel phantom was reported by Gibbs[8]. Zubal *et al* constructed an adult male phantom based on CT images from an adult male patient[9]. NORMAN and NAOMI using MRI data of healthy volunteers were introduced by Dimbylow[10]. In 2000, VIP-Man was developed by XU and his colleagues based on color photographic images[11].Many studies revealed that, due to simplified inequalities used to described the organs in MRID models, the inter-organ distances organ shape and location are variation with the real. For internal dosimetry, the influencing parameters are the relative position of source and target organs and organ mass. This leads to higher values of SAFs for many source–target organ combinations for the voxel models especially for lower photon energy[12]. Consequently, voxel phantoms could significantly contribute to improved dose assessment for patients. So, the International Commission on Radiological Protection(ICRP) and the International Commission of Radiation Units and Measurements(ICRU) decided to use the voxel-based models as a reference model to improve reference dosimetry[13] and the reference dosimetric parameters including SAFs are calculated using the new ICRP/ICRU adult reference phantoms[14].

*Supported by the Strategic Priority Research Program of Chinese Academy of Sciences (No. XDA03040000), National Natural Science Foundation of China (NO.910266004, NO.11305205, NO.11305203), the National Special Program for ITER (No. 2014GB112001)
E-mail: liqin.hu@fds.org.cn

Much tabulation of SAFs has been derived from Monte Carlo transport simulations using stylized computational models or voxel models to represent human internal organ anatomy. However, most of the phantoms are based on the medical images of Caucasian people may not completely appropriate for the application in the China. The internal radiation dose calculations built on Chinese voxel model become more and more important according to the development of nuclear medicine. Qiu R *et al* established the Chinese mathematical phantom to calculate the photon SAFs and compared with ORNL phantoms[15]. Liu Yang *et al* calculated the photon and electron SAFs based on the VCH voxel model by the color photographs on adult male cadaver[16]. However, within the investigations, there is a very limited works that reported using of voxel models represent the Chinese Female calculated SAFs for internal dosimetry.

In this study, a voxel model of Chinese adult female named Rad-HUMAN[17] was created using the color photographic images of the Chinese Visible Human (CVH) data set. A set of SAFs for monoenergetic photon and electrons were calculated using Rad-HUMAN which can represent the Chinese female. This present paper analyses the first set of SAFs calculated with Chinese female phantom by comparing the results with those of ICRP reference [14] and ORNL models. Dosimetric differences between mathematical and voxel models and those between the Chinese and Caucasian models will also be discussed in this paper.

2. Materials and methods

2.1 The Rad-HUMAN phantom

The high-resolution color photographic images of the CVH data set were obtained from a 22-year-old Chinese female cadaver. The candidate was 162 cm in height and 54 Kg in weight, which was close to the Chinese reference adult female[17]. The Rad-HUMAN which was shown in Fig.1 was constructed through three steps: 1. Identify and segment the organs and tissues in the color photographic images to yield 46 organs and tissues by experienced anatomists; 2. Assign the organs and tissues with density and chemical composition that recommended in ICRP 89 [18] and the International Commission on Radiation Units and Measurements (ICRU) 44 Report [19]; 3. Describe the anatomical data into Monte Carlo code input file[20]. After that, a voxel-based phantom that represented the average anatomical characteristics of the Chinese female population was established for radiation dosimetry.

However, manual description and verification of computational phantoms for MC simulation are tedious, error-prone and time-consuming. SuperMC/MCAM is a Multi-Physics Coupling Analysis Modeling Program[21-24] developed by the FDS Team[25-30]. Automatic conversion from CT/segmented sectioned images to human computational phantom can be performed by SuperMC/MCAM.

A whole-body computational phantom of Chinese female called Rad-HUMAN was created by SuperMC/MCAM using colored photographic images. Rad-HUMAN contains 46 organs and tissues and divided by more than 28.8 billion voxels, where 0.15mm x 0.15mm x 0.25mm voxel division for head and neck regions and 0.15mm x 0.15mm x 0.5mm for other regions.

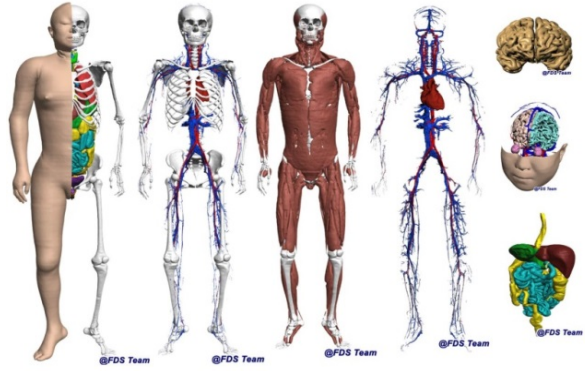

Figure 1. 3D view of Rad-HUMAN

## 2.2 Monte Carlo calculations

The Monte Carlo method has been widely used in situations where physical measurements and analytical clculations are either inconvenient or impossible. MCNP [31] is a general-purpose Monte Carlo code designed to transport neutrons, photons and electron in an arbitrarily assigned three-dimensional geometry. In this study, the Rad-HUMAN Phantom was implemented into the Monte Carlo particle transport code MCNP by repeated structure to describe the model. The density and element/chemical composition of organs and tissues acquired from ICRP 89 and ICRU Report 44 were used in the Monte Carlo simulations. According to the ORNL SAF data, mono-energetic and isotropic photon and electron sources were selected with discrete energy ranging from 10keV to 4MeV. Ten million electrons and photons histories were tracked per source region and energy in MCNP calculations.

MCNP offers a variety of variance reduction techniques based on different nonanalog simulations. It is important to use these techniques for the difficult problems to obtain both precise and computationally efficient results for solving difficult problems. It is difficult to obtain accurate SAFs when targets organ is small. There are three widely used variance reduction techniques to solve this problem: geometry splitting and Russian roulette, DXTRAN spheres and forced collisions (FCL) [31].The forced collisions (FCL) method which is more efficient than the other two methods was used in this simulation. The cutoff energy for both photon and electrons were set with the default values of 1keV.

## 2.3 Calculation method of SAFs

The SAF is an important quantity of organ self-dose and cross-dose for internal irradiation scenario. SAF is defined as the ratio of the fraction of energy emitted by radioactivity in a source organ ($r_S$) that is absorbed in the source organ itself, other target organs ($r_T$) and the target organ mass. According to the MIRD formalism [32], the equations related to absorbed dose, S values and the SAFs are as follws:

$$D(r_T, T_D) = \sum_{r_S} A(rs, T_D) \cdot S(r_T \leftarrow r_S) \quad (1)$$

where $D(r_T, T_D)$ is the absorbed dose in target. $A(rs, T_D)$ is the time-integrated or cumulated activity in source organ. And $S(r_T \leftarrow r_S)$ is the so-called S value in Gy·MBq$^{-1}$·S$^{-1}$ defined as the mean absorbed dose rate to target organ per nuclear transition in source tissue.

$$S(r_T \leftarrow r_S) = \sum_i E_i \, Y_i \, (rs, T_D) \cdot SAF(r_T \leftarrow r_S, E_i) \quad (2)$$

where $E_i$ is the mean energy of radiation type $i$. $Y_i$ is the yield of radiation type $i$ per transformation.

$$SAF(r_T \leftarrow r_S) = \frac{1}{m_T} \cdot \varphi_{i(t \leftarrow s)} = \frac{1}{m_T} \cdot \frac{E_T}{E_S} \quad (3)$$

where $m_T$ is the mass of the target. $\varphi_{i(t \leftarrow s)}$ is the absorbed fraction (AF). $E_T$ is the energy emitted from source organ and $E_s$ is the energy absorbed in target organs[12].

The electron SAFs calculated with Monte Carlo techniques for Rad-HUMAN phantom were compared with the former assumption of ICRP and MIRD [4,33] that electrons are fully absorbed in the source organ itself.

## 3. Results and discussion

The photon and electron SAFs were calculated using Rad-HUMAN phantom and compared to SAFs calculated using ORNL and ICRP/ICRU reference computational phantom. The SAFs for electrons were calculated using Rad-HUMAN and compared with SAFs from the assumptions of ICRP Publication[33].

### 3.1 Photon specfic absorbed fractions

Figure 2 showed the photon SAFs for self-irradiation. The source organ is also the target organ for the photon energy ranges from 10 keV to 4 MeV in many organs. The photon SAFs for self-irradiation decreases with increasing photon energy from 10keV-100keV. At photon energy of 0.1 MeV, the values begin to increase slightly to a maximum valule of 0.5 MeV and then begin to decrease again. Figure 2 showed the influence of organ mass on the photon SAFs for self-irradiation. It can be concluded that organs with small mass obtain larger SAFs than big mass organs, organs have the similar masses like kidney and stomach have very small differences in SFAs for self-irradiation.

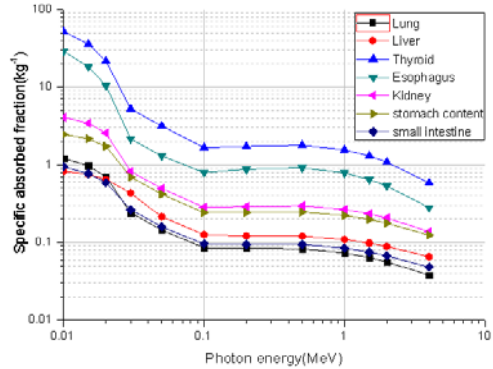

Figure 2. SAFs for photon self-absorption in some organs of Rad-HUMAN

Specific absorbed fractions (SAFs) for photon cross-absorption when liver as source were displayed in Figure 3. The figure showed that the SAFs of adrenal which has the small mass have the larger value, but the SAFs of esophagus and lung have the large mass and small differences in SAFs. Result shows that the organ geometry, density and the distance between source and target have the significant effect on the SAFs for cross-irradiation.

.

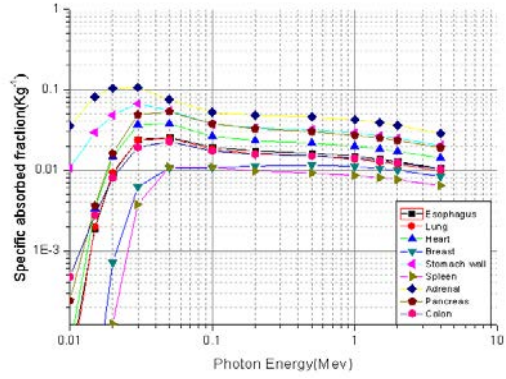

Figure 3. SAFs for photon cross-absorption as source in liver

The SAFs using Rad-HUMAN phantom were compared with SAFs from ORNL and ICRP adult reference phantoms. Figure 4 shows the SAFs for the photon self-absorption in lungs of the two voxel phantoms and the mathematical phantom. The SAFs for organs in Rad-HUMAN Phantom have the similar tendency with those in the ICRP/ICRU reference phantom and ORNL phantom. The SAFs values for photon self-absorption in lung were larger than the other models for has less mass. Variations in SAFs were calculated as the ratio of the Rad-HUMAN to the ORNL and ICRP adult reference phantoms. The average ratio was 20% with the ICRP adult reference phantoms and 48% with the ONRL phantom. In case of the self-irradiation, the variations are dependent on the difference organ mass; the organ geometry does not have a significant influence on the SAF estimation. Figure 5 showed that the SAFs (lung->stomach wall) using the Rad-HUMAN diverge from the other models. For energy bellow 0.5 MeV the ratio reached up to 50% and for higher energy the ratio decreased to 20%. The observed discrepancies are due to different shapes and inter-organ distances between the organs of the phantoms whose influence is quite dominant at low energy.

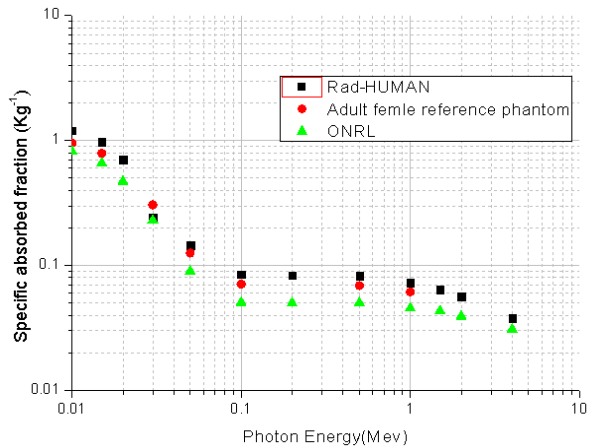

Figure 4.SAFs for photon self-absorption in lung

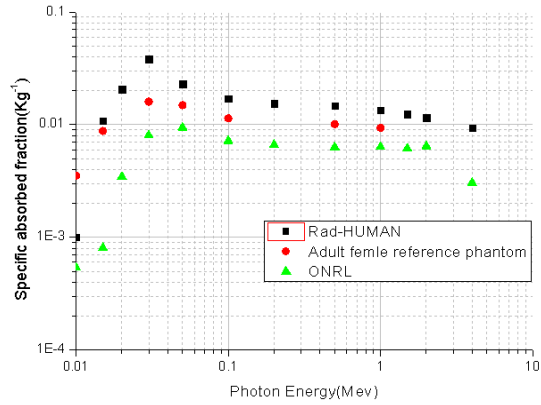

Figure 5. SAFs for photon cross-absorption (lung -> stomach wall)

**3.2 Electron specific absorbed fractions**

Because of low penetration power of electrons and the previously applied assumption of ICRP Publication 30[33] which supposed the electrons are fully absorbed in the source organ and electron, AFs are recommended to be 1, AFs and SAFS are recommended to be 0 when the source and the target are different.

Figure 6 shows the electron SAF values for self-absorption in many organs of the Rad-HUMAN phantom. The SAF values for electrons are different to the simplified assumptions of ICRP Publication 30. From Figure 6 we can see that electrons have the ability to leave the source organ with electron energy above 0.5MeV. The self-absorption SAFs are constant and agree with the inverse organ mass for electron energy. For large organs such as liver and lung, the drop-off of the AFs and SAFs with increasing electron energy is moderate as electron energy increases, since short electron ranges still in the large source regions. For small organs like thyroid, the drop-off of the AFs and SAFs are much more pronounced with the increase of electron energy, since even shorter electron ranges could cross the organ boundary.

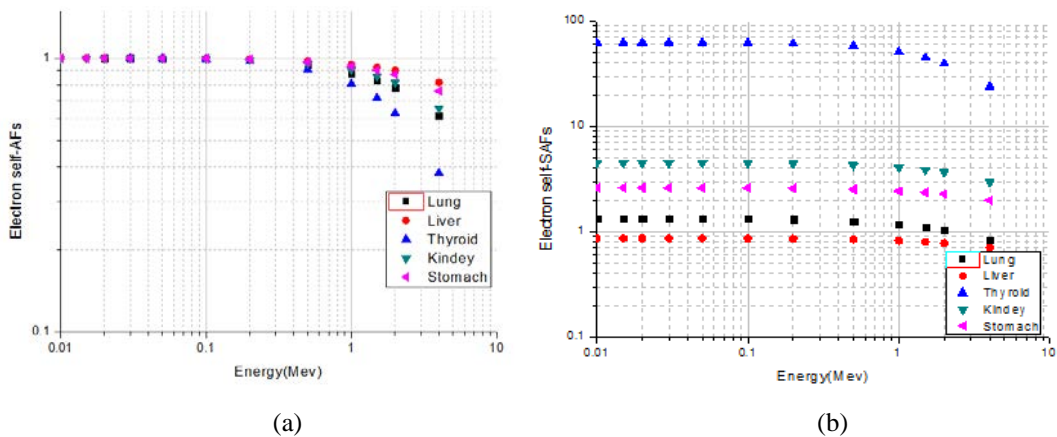

(a)                                   (b)

Figure 6. AFs(a)and SAFs(b) for Electron Self-absorption

Figure 7 shows the cross-absorption electron SAF values for source in stomach content of the Rad-HUMAN phantom. Figure 8 shows the electron irradiation of adjacent regions cannot be always neglected, even though electrons are considered as weak penetrating radiation. Results show that SAF values for distant organ such as thyroid are smaller than other organs because of short electron ranges. The values for neighboring organs such as spleen and liver cannot be negligible for electron energy above 1MeV.

From Monte Carlo calculation of electron SAF values, we can conclude that the high-energy electrons can cross the source organs boundary and the ICRP 30 approach assuming full absorption in the source underestimates the absorption of the neighboring organs around the source organs. The SAF value is related to the geometry and distance between the source and organ. The real phantom and Monte Carlo transport method could make the dosimetry calculation clinically possible in nuclear medicine.

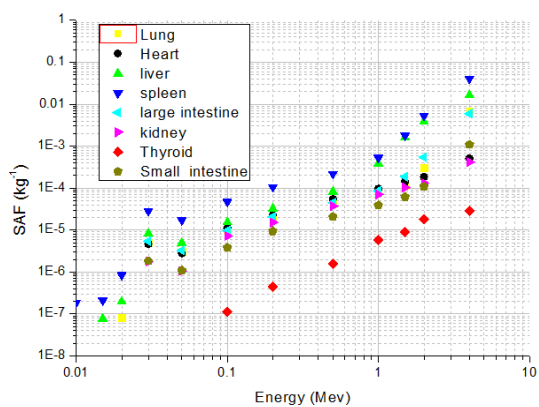

Figure 7. Electron SAFs for source in stomach content

### 3.3 S value calculation

Once mono-energetic photon and electron SAFs are assembled, S values for γ-and β-ray emitted from radionuclides were calculated using Eq.2. In this study, we calculated the radionuclides associated with common molecular studies of the $^{99m}$Tc [34] which usually were used as molecular imaging radionuclides. S values of $^{99m}$Tc were calculated in the liver and compared with those of VIP-Man, VCH and MIRD as listed in Table 1[16].

The radionuclide $^{99m}$Tc are only gamma emitters. From the table we can see significant variation in S values between those phantoms. The S Value of adrenal of Rad-Human is 62% larger than the S Value of adrenal of VCH, 30% larger than the S Value of adrenal of VIP-Man and 180% larger than the S Value of adrenal of MIRD Pamphlet No.11. The values' differences among those phantoms were due to variations in organ mass, organs size and organ contours. Small organs such as adrenal and pancreas are difficult to segment and have the irregular shape. The organ of those phantoms may have large differences in size and mass. Differences of S value for $^{99m}$TC in liver to other model S values were primarily due to variations in organ size, volume, mass and inter-organ separation.

**Table 1** Comparisons of S-values and mean absorbed doses for organs with $^{99m}$Tc distributed in liver with VCH, VIP-Man and MRID Pamphlet N0.11

| Organs | Rad-HUMAN | VCH | VIP-Man | MIRD Pamphlet No.11 |
|---|---|---|---|---|
| Adrenal | $9.79\times10^{-16}$ | $6.01\times10^{-16}$ | $7.48\times10^{-16}$ | $3.38\times10^{-16}$ |
| Kidney | $3.87\times10^{-16}$ | $2.54\times10^{-16}$ | $3.65\times10^{-16}$ | $2.93\times10^{-16}$ |
| Liver | $3.56\times10^{-15}$ | $3.49\times10^{-15}$ | $3.36\times10^{-15}$ | $3.45\times10^{-15}$ |
| Lungs | $3.62\times10^{-16}$ | $7.77\times10^{-16}$ | $2.80\times10^{-16}$ | $1.88\times10^{-16}$ |
| Pancreas | $3.46\times10^{-16}$ | $3.38\times10^{-16}$ | $5.50\times10^{-16}$ | $3.15\times10^{-16}$ |
| Spleen | $2.17\times10^{-16}$ | $1.44\times10^{-16}$ | $1.27\times10^{-16}$ | $6.91\times10^{-16}$ |
| Stomach | $7.49\times10^{-16}$ | $4.59\times10^{-16}$ | $4.98\times10^{-16}$ | $1.43\times10^{-16}$ |
| Thyroid | $4.17\times10^{-17}$ | $8.57\times10^{-17}$ | $5.06\times10^{-17}$ | $1.13\times10^{-17}$ |

## 4. Conclusion

In this study, a new set of SAFs and S values of the Rad-HUMAN were calculated and compared with the SAF data of ORNL and ICRP references phantom. The first set of SAFs using Rad-HUMAN which can represent the Chinese female could make the dosimetry calculation more exact in nuclear medicine for Chinese female. In the present study, it has been confirmed that the SAFs for self-irradiation depended on the energy and the mass of the target/source organ, the SAFs for cross irradiation depended on the relative position of source and target organs. It can be concluded that SAF for Rad-HUMAN have the similar trends that validate the data is accurate and reliable for internal radiation dose calculations for Chinese female. The SAFs and S values obtained using the real phantom and connected to the individual biokinetic data could make the dosimetry calculation clinically possible in nuclear medicine.

*We would like to express our appreciation for the support of the members of FDS Team. We would like to thank Professor Shaoxiang Zhang et al, from the Third Military Medical University, People's Republic of China, for providing the visible human anatomical dataset and related support, and Professor X. George Xu for helpful discussion and support during the work.*

**REFERENCES**

1. Loevinger R., Berman M. MIRD J Nuclear Med 1968; Suppl 1.
2. Loevinger R, Berman M. NM/MIRD Pamphlet no. 1, Rev., March 1976, Society of Nuclear Medicine, NY, 1976.
3. Snyder WS, Fisher Jr, Ford MR, et al. Nucl Med,1969,3:7-52.
4. Snyder WS, Ford MR, Warner GG. The Society of Nuclear Medicine, 1978.
5. Cristy M. Report ORNL/NUREG/TM-367, Oak Ridge National Laboratory, Oak Ridge, Tenn., USA, 1980.
6. Kramer R, Zankl M, Williams G. GSF - Report S-885.Reprint July 1999.Institut für Strahlenschutz, GSF-Forschungszentrum für Umwelt und Gesundheit, Neuherberg-München, 1982.
7. Stabin M, Watson E, Cristy M, Ryman J, Eckerman K, Davis J, Marshall D. and Gehlen K. Report No. ORNL/TM-12907, Oak Ridge National Laboratory, Oak Ridge, Tenn., USA, 1995.
8. Gibbs SJ, Pujol A Jr, Chen TS, et al. Oral Surg Oral Med O, 1984, 58: 347–354.
9. Zubal I G, Harrell C R, Smith E O, Rattner Z, Gindi G and Hoffer P B. Med Phys.1994,21 299-302
10. Dimbylow P J.Phys.Med.Biol,2005,50 1047-70.
11. Xu XG,Chao TC,Bozkurt A. Health Phys,2000,78（5）: 476-485．
12. Zankl M,Schlattl H,Petoussi-Henss N, et al. Voxel Phantoms for internal Dosimetry[M]/Radiation Physics for Nuclear Medicine. Springer Berlin Heidelberg,2011:257-279.
13. ICRP 2007 Recommendations of the International Commission on Radiological Protection ICRP Publication 103 (Oxford: Elsevier)
14. L Hadid, A Desbree, H Schlattl, D Franck, E Blanchardon, M Zankl, Phys. Med. Biol. 2010,55,3631-3641
15. Qiu R, Li J L, Zhang Z, et al. Health Phys, 2008, 95:716-724.
16. LIU Yang, XIE Tianwu, LIU Qian, Nuclear Science and Techniques, 2011,22,165-173.

17. CHENG Mengyun. Computational Modeling Method and Its Application[D].Hefei: Chinese Academy of Science,2012.
18. International Commission on Radiological Protection, Basic Anatomical and Physiological Data for Use in Radiological Protection: Reference Values, ICRP Publication 89, 2003.
19. International Commission on Radiation Units and Measurements, Tissue substitutes in radiation dosimetry and measurement, Bethesda, MD: ICRU; ICRU Report 44, 1989.
20. Mengyun Cheng, Qin Zeng, Ruifen Cao, et al. Progress in Nuclear Science and Technology (PNST), 2011, 2: 237-241.
21. Y. Wu, FDS Team.Fusion Engineering and Design, 2009,84,1987-1992
22. Y.WU,Z.Xie,U.Fischer.Nuclear Scinece and Engineering and Design,2009,84(7-11),1987-1992
23. Y. Li, L. Lu, A. Ding, etal. Fusion Engineering and Design. 2007, 82, 2861-2866.
24. Y.Wu, J. Song, H.Zheng, et al.Annals of Nuclear Energy.DOI:10.1016/j.anucene.2014.08.058
25. Y. C. Wu, J. Q. Jiang, M. H. Wang, M. Jin, and FDS Team,, Nuclear Fusion, 2011, 51(10) 103036.
26. Y. Wu, H.chen, S. Liu, et al, Journal of Nuclear Materials, 2009, 386-388:122-126.
27. Y. Wu, FDS Team, Fusion Engineering and Design, 2008, 83:1683-1689.
28. Y. Wu, FDS Team. Fusion Engineering and Design, 2006, 81: 2713-2718.
29. Y. C. Wu, J. P. Qian, J.N Yu. Journal of Nuclear Materials, 2002, vol. 307-311: 1629-1636.
30. WU Yi-Can, SONG Gang, CAO Rui-Fen, et al, Chinese Physics C (HEP & NP), 2008, 32(Suppl. II): 177-182.
31. X-5 Monte Carlo Team. MCNP5-A General Monte Carlo N-Particle Transport Code, Volume II: Users Guide[R]. New Mexico: Los Alamos National Library, 2003.
32. Bolch WE, Eckerman KF, Sgouros G, Thomas SR. J Nucl Med. 2009; 50:477-484. [PubMed: 19258258]
33. ICRP 1979 Limits for intakes of radionuclides by workers: part 1 ICRP Publication 30 (Oxford: Pergamon)
34. ICRP. ICRP Publication 107: Nuclear decay data for dosimetric calclations. Ann ICRP. 2008b; 38:1–26. [PubMed: 19154964]